\documentclass{emulateapj}
\usepackage{apjfonts}
\usepackage{graphicx}

\begin{document}

\title{Photospheric emission as the dominant radiation mechanism
  in long-duration gamma-ray bursts}

\shorttitle{GRB photospheres}

\author{Davide Lazzati\altaffilmark{1}, Brian
  J. Morsony\altaffilmark{2}, Raffaella Margutti\altaffilmark{3}, and
  Mitchell C. Begelman\altaffilmark{4,5}} \shortauthors{Lazzati et
  al.}

\altaffiltext{1}{Department of Physics, NC State University, 2401
  Stinson Drive, Raleigh, NC 27695-8202}

\altaffiltext{2}{Department of Astronomy, University of
  Wisconsin-Madison, 3321 Sterling Hall, 475 N. Charter Street,
  Madison WI 53706-1582}

\altaffiltext{3}{Harvard-Smithsonian Center for Astrophysics, ITC, 60
  Garden Street, Cambridge, MA 02138, USA}

\altaffiltext{4}{JILA, University of Colorado and National Institute
  of Standards and Technology, Boulder, CO 80309-0440, USA}

\altaffiltext{5}{Department of Astrophysical and Planetary Sciences,
  University of Colorado, Boulder, CO 80309-0391, USA}

\begin{abstract} 
  We present the results of a set of numerical simulations of
  long-duration gamma-ray burst jets associated with massive, compact
  stellar progenitors. The simulations extend to large radii and allow
  us to locate the region in which the peak frequency of the advected
  radiation is set before the radiation is released at the
  photosphere. Light curves and spectra are calculated for different
  viewing angles as well as different progenitor structures and jet
  properties. We find that the radiation released at the photosphere
  of matter-dominated jets is able to reproduce the observed Amati and
  energy-Lorentz factor correlations. Our simulations also predict a
  correlation between the burst energy and the radiative efficiency of
  the prompt phase, consistent with observations.
\end{abstract}


\section{Introduction}

Long-duration gamma-ray bursts (GRBs) are characterized by bright,
non-thermal radiation spectra with typical photon energies of a few
hundred keV to several MeV (Band et al 1993; Kaneko et al. 2006; Zhang
et al. 2011). The standard model to explain these observations invokes
collisionless shocks between parts of the relativistic outflows moving
at different speed (Rees \& Meszaros 1994) to produce a tangled
magnetic field and non-thermal electrons, eventually releasing
synchrotron radiation (Daigne \& Mochkovitch 1998; Piran 1999,
Bo\v{s}njak et al. 2009). Alternatively, the magnetic field may be
advected from the central engine and the collisions working as a
trigger for magnetic dissipation (the ICMART model, Zhang \& Yan
2011).

When GRB observations are corrected for redshift and considered in the
burst's frame, a correlation between the burst's isotropic equivalent
energy and its characteristic photon energy is revealed (Amati et
al. 2002). The correlation was originally discovered with data from
BeppoSAX, and dubbed the Amati correlation. Even though the role of
selection effects has not been completely understood (Nakar \& Piran
2005; Kocevski 2012), the correlation has been confirmed with data
from all subsequent observatories (Amati 2006; Ghirlanda et al. 2008;
Amati et al. 2009). To lend credibility to an intrinsic origin, a
correlation between burst peak luminosity and peak photon energy is
also observed among bursts (Yonetoku et al. 2004; Ghirlanda et
al. 2012a) and within individual events (Lu et al. 2012). The latter
correlation is free of selection effects and points to a physical
origin for all the others. More recently, the analysis of a sample of
GRBs with very early afterglow observations revealed a correlation
between the isotropic equivalent energy of the prompt emission and the
Lorentz factor of the outflow (Liang et al. 2010; Ghirlanda et
al. 2012b). Taken together, these correlations present a challenge to
the standard synchrotron shock model (SSM).  Outflows characterized by
large Lorentz factors should produce internal shocks at large radii,
therefore generating magnetic fields of much lower intensity, and
eventually bursts characterized by smaller photon energies, predicting
an anti-correlation in contrast to observations (Ghisellini et
al. 2000).

An alternative model for the prompt emission of GRBs is the
photospheric model (M\'esz\'aros \& Rees 2000; Pe'er et al. 2005,
2006; Rees \& M\'esz\'aros 2005; Giannios 2006; Lazzati et
al. 2009). It does not specify how the photons are produced,
concentrating instead on how the interaction of the radiation field
with the leptonic component of the outflow modifies the spectrum in
the optically thick phase, before it is released at the
photosphere. The photospheric model has been proposed as a viable
candidate to reproduce correlations such as the Amati correlation on
the qualitative level (Thompson et al. 2007; Lazzati et al. 2011; Fan
et al. 2012). On the quantitative side, comparison of the burst
properties from numerical simulations to observations have shown that
while photospheric emission from baryonic jets can reproduce the slope
of the correlation, they cannot reproduce the normalization, simulated
bursts being more energetic than observed ones for a given peak
frequency (Lazzati et al. 2011; Nagakura et al. 2011). Deeper
theoretical understanding of photospheric emission has revealed that
radiation and matter are not in equilibrium all the way to the
photosphere and therefore the peak of the spectrum is formed at
moderate optical depths ($\tau\sim50$) when the radiation is at higher
temperature (Giannios 2012).

In this paper we present results of an extensive set of simulations
aimed at exploring quantitatively the predictions of the photospheric
model for baryonic jets. In contrast to our previous work, the peak
frequency of the spectrum is calculated at an optical depth larger
than unity, following Giannios (2012). We also explore the effect of
the jet injection properties and of the progenitor structure on the
ensuing light curves and spectra. This paper is organized as follows:
in Sect. 2 we present our numerical simulations, in Sect. 3 we detail
the post processing methods to derive light curves and spectra, in
Sect. 4 we present our results, and in Sect. 5 we summarize and
discuss this work.

\section{Numerical simulations}

All the simulations presented in this paper were performed with the
FLASH code (Fryxell et al. 2000), version 2.5. The FLASH code is a
modular block-structured adaptive mesh refinement code, parallelized
using the Message Passing Interface (MPI) library. It solves the
Riemann problem using the Piecewise Parabolic Method (PPM). From the
point of view of the research presented here, the main strength of
this code is that it can perform special relativistic hydrodynamic
computations on an adaptive mesh. This is crucial for the modeling of
the interaction of the collimated jet with the stellar envelope. This
approach allows for significantly enhanced resolution along the jet
axis and the jet-star boundary without the need to resolve the rest of
the star to the same extent, thereby making high-resolution
simulations feasible.  In order to increase the resolution near the
center of the star and close to the axis of the jet we modified FLASH
to be able to vary the maximum level of refinement allowed over the
simulation grid (see also L\'opez-C\'amara et al. 2013).  Additional
modifications to the standard FLASH code are described in Morsony et
al. (2007).

We adopted a maximum resolution of $4\times10^6$~cm at the highest
level of refinement. At this resolution the transverse dimension of
the injected jet is resolved into 44 elements. Our simulations do not
include magnetic fields, due to the technical challenge of performing
MHD calculations with relativistic motions on an adaptive mesh. In
addition, gravity from a central mass and self-gravity are neglected
since the characteristic times of the jet-star interaction are much
shorter than the dynamical time of the progenitor star's collapse.

\begin{figure}
\includegraphics[width=\columnwidth]{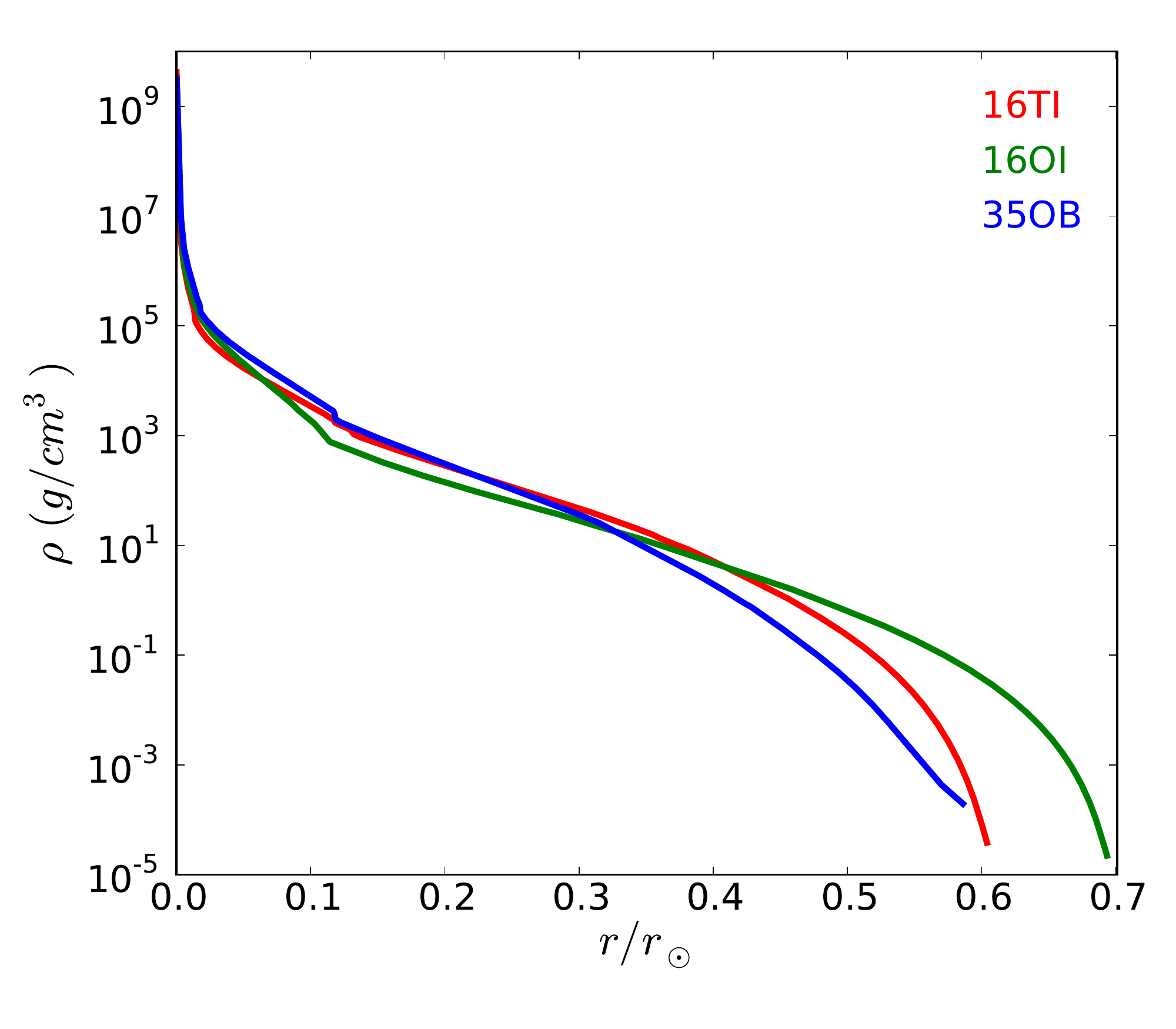}
\caption{{Density profiles of the GRB progenitor stars used in this
    work.}
\label{fig:proge}}
\end{figure}

All our simulations adopted a realistic GRB stellar progenitor, and we
explored three different progenitors all taken from Woosley \& Heger
(2006). Model 16TI is a 16 solar-mass Wolf-Rayet star with an initial
metallicity 1\% solar and angular momentum
$J=3.3\times10^{52}$~erg$\cdot$s. The mass of the star at
pre-explosion is 13.95 solar masses and its radius is
$4.07\times10^{10}$~cm, corresponding to 0.6 solar radii. A
relativistic jet is injected as a boundary condition at a distance
$R_0=10^9$~cm from the star's center. The jet has initial Lorentz
factor $\Gamma_0=5$ and an half-opening angle $\theta_0=10^\circ$
($5^\circ$ in one case). The jet is injected hot, with enough internal
energy to allow for acceleration up to an asymptotic Lorentz factor
$\Gamma_\infty=400$ upon complete, non-dissipative acceleration
($\Gamma_\infty=100$ upper limit was adopted in one case). The engine
luminosity is kept constant for a duration $t_{\rm{eng}}$ (typically
$t_{\rm{eng}}=100$~s, but shorter engines were explored as well),
after which the jet is turned off and the boundary condition is set to
reflective. The bottom and polar boundaries are set to reflective for
the whole duration of the simulation, while the two outer boundaries
are set to absorbing. The simulation box is $2.5\times10^{13}$~cm in
length (along the jet direction) and $5\times10^{12}$~cm across,
approximately a factor 10 larger than our previous simulations.

Two additional progenitor stars were tested. Model 16OI is a star of
16 solar masses with an initial metallicity 10\% solar. Its
pre-explosion mass is 12.21 solar masses and its radius is 0.7 solar
radii. Model 35OB is a 35 solar-mass star with initial metallicity
10\% solar, pre-explosion mass 21.24 solar masses, and radius 0.6
solar radii. The density profiles of the three progenitors are shown
in Figure~\ref{fig:proge}, while details of the simulations are
reported in Table~\ref{tab:sim}.

\begin{figure}
\includegraphics[width=\columnwidth]{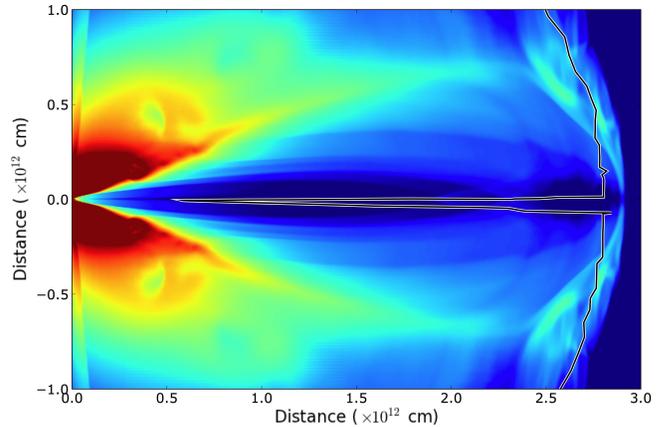}
\caption{{False-color stratification map of the logarithm of the
    density of our fiducial 16TI simulation at $t=100$~s. The thick
    black line shows the location of the photosphere for an observer
    located at a viewing angle $\theta_o=0.5^\circ$ towards the bottom
    of the panel.}
\label{fig:frame}}
\end{figure}

\begin{table}[t] 
\bigskip
\begin{center}
\begin{tabular}{c|c|c|c|c|c|c}
  Sim. \# & Prog. ID & $\Gamma_0$ & $\Gamma_\infty$ & $\theta_0$ &
  $L_{\rm{jet}}$ (erg/s) & $t_{\rm{eng}}$ (s)\\
  \hline
  1 & 16TI & 5 & 400 & 10 & $5.33\times10^{50}$ & 100 \\
  2 & 16TI & 5 & 100 & 10 & $10^{50}$ & 100 \\
  3 & 35OB & 5 & 400 & 10 & $5.33\times10^{50}$ & 100 \\
  4 & 16OI & 5 & 400 & 10 & $5.33\times10^{50}$ & 100 \\
  5 & 16TI & 5 & 400 & 10 & $10^{50}$ & 100 \\
  6 & 16TI & 5 & 400 & 5 & $5.33\times10^{50}$ & 100 \\
  7 & 16TI & 5 & 400 & 10 & $5.33\times10^{50}$ & 67 \\
  8 & 16TI & 5 & 400 & 10 & $5.33\times10^{50}$ & 30 
\end{tabular}
\end{center}
\caption{{Details of the simulations presented in this work.}
\label{tab:sim}}
\end{table}

\section{Spectra and Light Curves}
 
The calculation of the light curves and spectra is performed
analogously to Lazzati et al. (2011, see also Mizuta et al. 2011;
Nagakura et al. 2011) with a few important changes. First, we changed
the optical depth at which we compute the radiation temperature. In
our previous publications, we used the photospheric radius as the
location at which the radiation temperature was calculated through the
black body relation $u=aT^4$, where $u$ is the energy density and
$a=7.56\times10^{-15}$~erg~cm$^{-3}$~K$^{-4}$ is the radiation energy
constant. We now use Eq. 6 from Giannios (2012) to derive the opacity
at which the radiation temperature is set:
\begin{equation}
\tau_T=46\frac{L_{53}^{1/6}f_\pm^{1/3}}{\Gamma_{2.5}^{1/3}\epsilon^{1/6}\eta_{2.5}^{1/3}}
\end{equation} 
where $f_\pm$ is the number of leptons per proton (including
electron-positron pairs), $\epsilon$ is the fraction of the energy of
the outflow carried by radiation, $\eta=L/\dot{M}c^2$ is the
asymptotic Lorentz factor, and we used the notation
$Q_x=10^{-x}Q$. Because of the decrease in the radius with respect to
the standard assumption $\tau=1$, the location at which the spectrum
is formed is easier to determine from the simulation frames and no
correction needs to be applied (see Eq. 4 in Lazzati et al. 2011). Our
numerical simulations do not allow us to compute the number of
electron-positron pairs and we assume that $f_\pm=1$ for the remainder
of this paper. This assumption is justified by the fact that the
comoving temperature is of only a few keV or less where the peak
frequency is computed and, in absence of non-thermal radiation, pairs
are not important. If a significant population of pairs were present,
the location of the spectrum formation would be moved outwards and the
peak frequency would be decreased, weakening the agreement with
observations (see figures and discussion below).

The location of the region of the spectral peak formation was
calculated analogously to Lazzati et al. (2011) but updating their
condition to:
\begin{eqnarray}
&&46\frac{L_{53}^{1/6}f_\pm^{1/3}}{\Gamma_{2.5}^{1/3}\epsilon^{1/6}\eta_{2.5}^{1/3}}=
\nonumber \\ 
&&=-\int_{Z_{\rm{obs}}}^{Z_{\rm{ph}}(x)} 
\sigma_T n^\prime\left(t_{\rm{obs}}-\frac{Z_{\rm{obs}}-z}{c},x,z\right)\,
\Gamma\left[1-\beta\cos(\theta_v)\right]\,dz
\end{eqnarray}
where $\beta\equiv\beta(t_{\rm{lab}},x,z)$ is the local velocity of
the outflow in units of the speed of light,
$\Gamma\equiv\Gamma(t_{\rm{lab}},x,z)$ is the local bulk Lorentz
factor, and $\theta_v\equiv\theta_v(t_{\rm{lab}},x,z)$ is the angle
between the velocity vector and the direction of the line of
sight. $x$ is the coordinate perpendicular to the line of sight, while
$z$ is the coordinate along the line of sight.
All the values of $\beta$, $\Gamma$, and $\theta_v$ are
evaluated at the same delayed coordinate
$(t_{\rm{lab}},x,z)\equiv\left(t_{\rm{obs}}-\frac{Z_{\rm{obs}}-z}{c},x,z\right)$
as the comoving density.

Figure~\ref{fig:frame} shows an image of the density stratification
from our fiducial simulation (simulation number 1 in
Table~\ref{tab:sim}) at a time $t=100$~s after the engine onset. A
black line is used to show the location of the photosphere for an
observer located at a viewing angle $\theta_o=0.5^\circ$ towards the
bottom of the panel. Most of the photospheric emission comes from the
bottom of the trough where the outflow velocity vector and the
radiation propagation are aligned within an angle $\sim1/\Gamma$.

\begin{figure}
\includegraphics[width=\columnwidth]{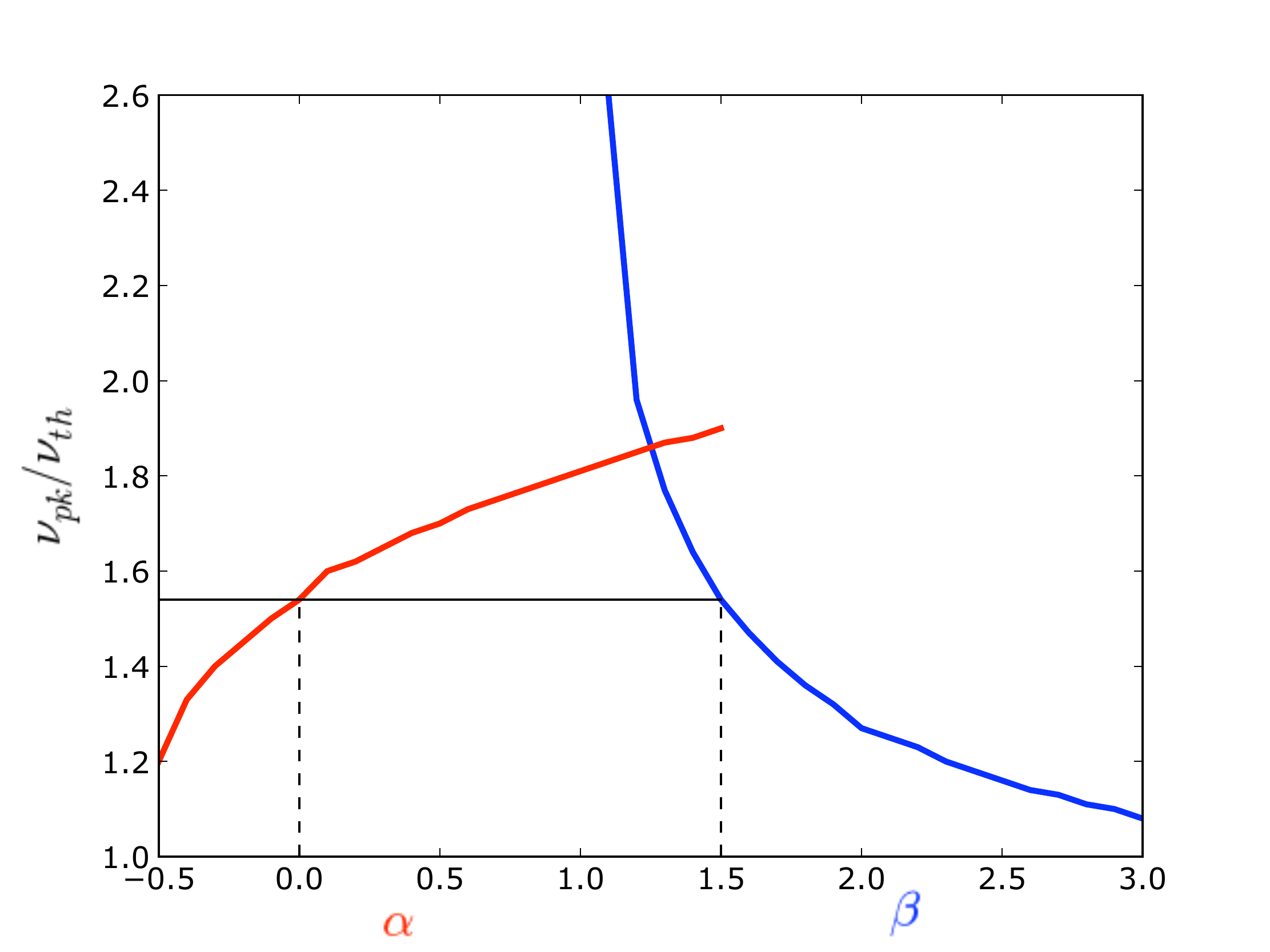}
\caption{{Modification of the location of the $\nu F(\nu)$ peak of a
    Planck spectrum due to the addition of a Comptonized
    high-frequency tail (blue) or low-frequency tail (red). Indices in
    the x-axis are spectral indices ($F(\nu)\propto\nu^{-\alpha}$;
    $F(\nu)\propto\nu^{-\beta}$).}
\label{fig:nupk}}
\end{figure}

In addition, we refined the way in which we compute the peak frequency
to make the comparison with observations more accurate. First, we
consider the fact that what is measured from the data is the peak of
the $\nu F(\nu)$ spectrum. Second, we consider the fact that the
location of the peak frequency is affected by the addition of
high-frequency power-law tails likely due to Comptonization (Pe'er et
al. 2006; Giannios \& Spruit 2007; Beloborodov 2010; Lazzati \&
Begelman 2010) and low-frequency power-laws likely due to a
synchrotron component (Vurm et al. 2011). The combination of these two
effects shifts the peak to higher frequencies by approximately a
factor 2, depending on the slope of the high- and low- frequency
power-law components (see Figure~\ref{fig:nupk}).

\begin{figure}
\includegraphics[width=\columnwidth]{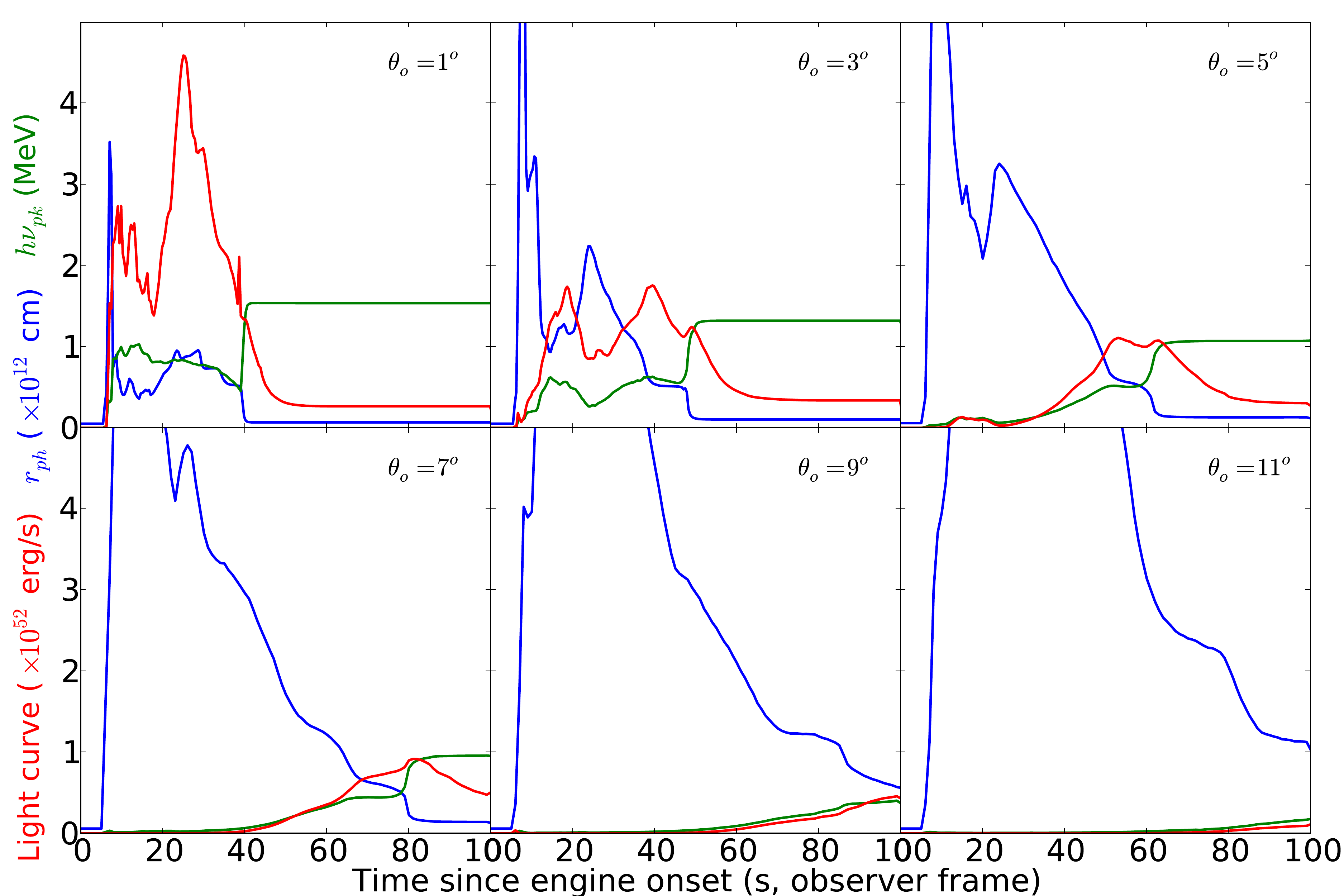}
\caption{{Sample results of our post processing for simulation \#1
    (see Table~\ref{tab:sim}). Each panel shows a different
    viewer. Different colors show the bolometric light curve (red),
    the radius of the photosphere (blue) and the peak frequency
    (green).}
\label{fig:lc}}
\end{figure}

A sample of light curves, photospheric radii, and peak frequencies are
plotted vs. time in Figure~\ref{fig:lc}. These light curves are quite
representative also of the other simulations. As a matter of fact, the
effects of the viewing angle are much more pronounced than those of a
different progenitor, engine duration, or jet properties. As we move
away from the jet symmetry axis, all bursts grow weaker,
start at a later time, are released at a larger photospheric radius,
and are characterized by a smaller photon frequency.

\section{Correlations}

\begin{figure}
\includegraphics[width=\columnwidth]{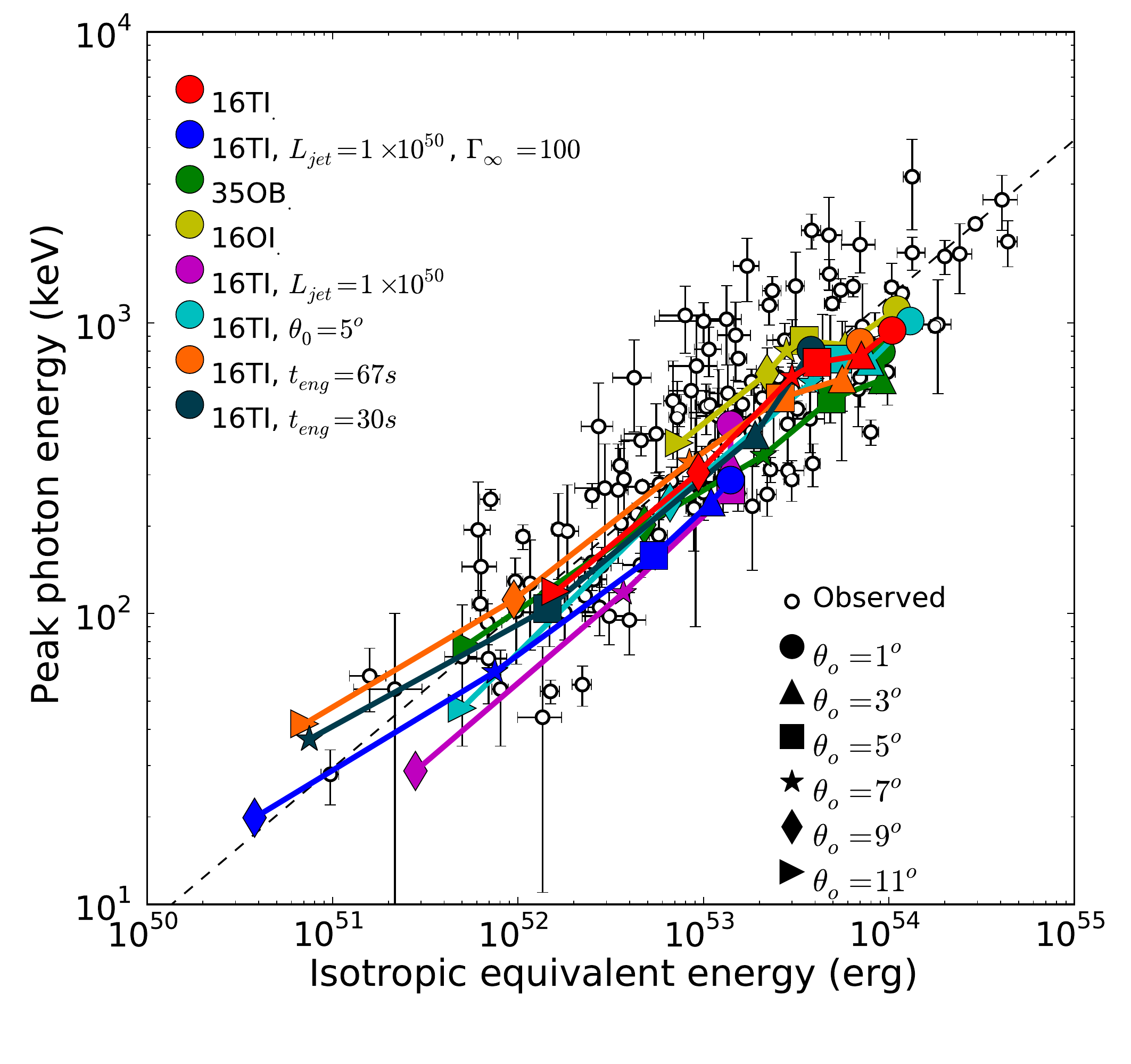}
\caption{{Overlay of the simulation results on the observed spectral
    and energetic properties of gamma-ray bursts.  The black symbols
    with error bars show the observed Amati correlation, i.e., the
    correlation between the isotropic-equivalent energies of the
    bursts and the peak photon energies of their $\nu F(\nu)$ spectra
    (Amati et al. 2002; 2009).  The colored lines and symbols show the
    results of our simulations.  Different colors refer to different
    sets of progenitor and jet properties, as described in the upper
    left legend.  The legend states explicitly the parameters of the
    simulation that differ from our fiducial progenitor/jet pair (see
    text and Table~\ref{tab:sim}).  Different symbols refer to
    different orientations of the observer with respect to the jet
    axis, as described in the lower right legend.}
\label{fig:amati}}
\end{figure}

Figure~\ref{fig:amati} shows the observed Amati correlation (Amati et
al.  2002; 2009) compared to the one found from our synthetic light
curves. The change in the location of the radius at which the spectrum
is formed has great influence on the synthetic light curves. The
typical photon frequency is increased while the total energy is almost
unchanged. As a consequence, synthetic bursts agree quantitatively
with the Amati correlation in both the slope and the
normalization. The agreement results from the combination of two
effects. First, the shearing and shocking effects of the stellar
material on the outflowing plasma cause a polar stratification of the
jet. Different observers see different bursts that, when placed on the
Amati diagram, form a stripe along the correlation. In addition, the
pressure exerted by the stellar material on the jet walls collimates
the outflow in such a way as to produce large-scale jets that are
insensitive to their initial conditions. We also find that the most
important parameter to determine the location of a burst on the Amati
diagram is the radius of its stellar progenitor: the slightly bigger
16OI progenitor produces higher peak frequencies at a given energy
than the more compact 16TI and 35OM progenitors.

The agreement between the synthetic bursts and the observed
correlation is quite remarkable since the photospheric model has only
a few free parameters: the progenitor star structure, the outflow
luminosity and its asymptotic Lorentz factor, and the opening angle of
the injected jet. We tried different values for each of these
parameters and we always found complete agreement between the
simulated and observed bursts. We also tried bursts from engines with
different durations. As long as the jet reaches the star surface still
relativistic, the computed bursts lie on the Amati correlation. We
finally note that our simulations do not cover extreme bursts with
very high luminosity and Lorentz factors but are rather targeted to
reproduce ``average'' events. For that reason, none of our simulations
reproduces the data in the upper right corner of the figure.

\begin{figure}
\includegraphics[width=\columnwidth]{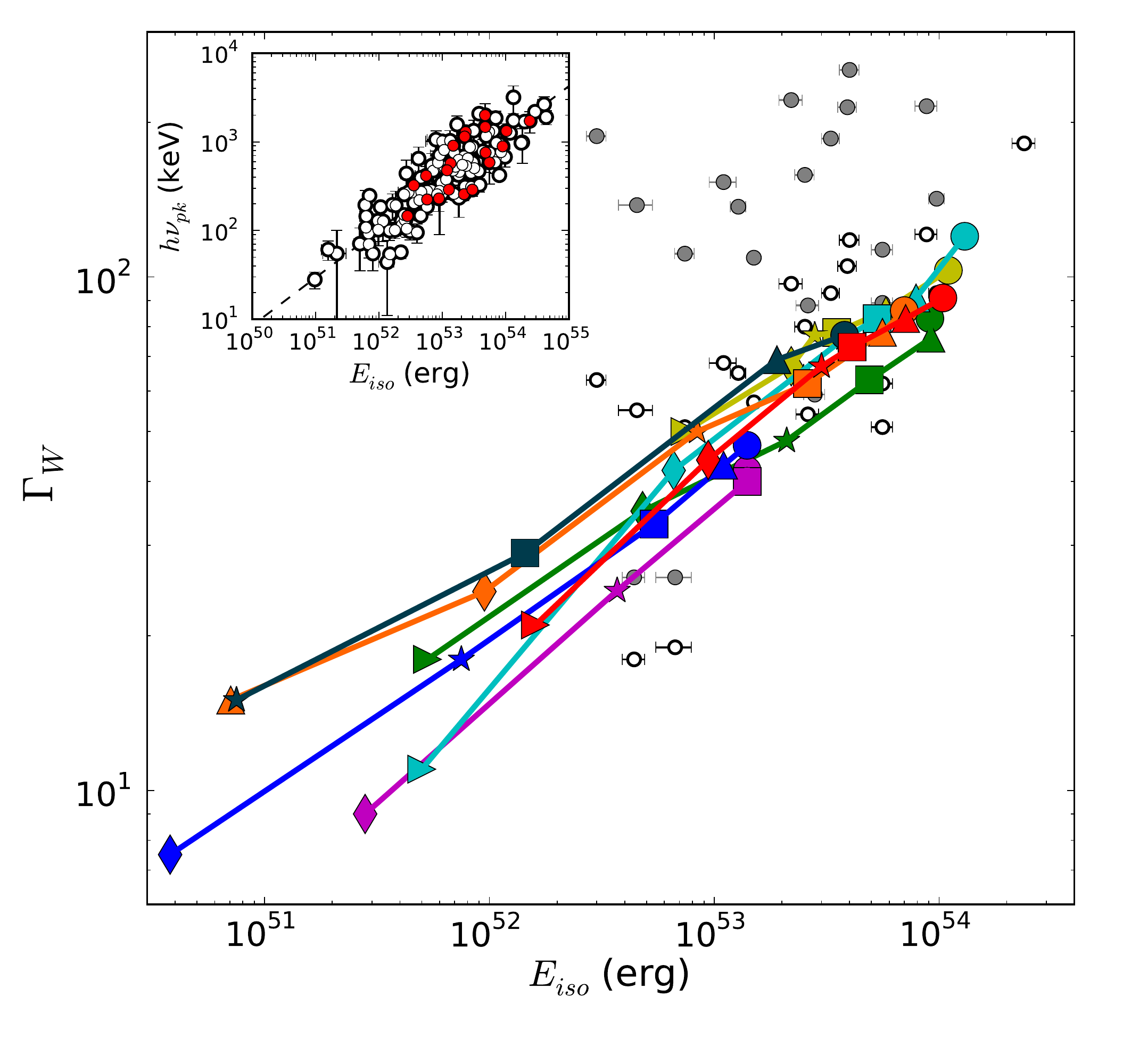}
\caption{{Comparison between observed and simulated outflow Lorentz
    factor versus the burst energy.  The black symbols with error bars
    show the measured Lorentz factor versus the isotropic equivalent
    energy, assuming that the external shock propagates into a wind
    environment.  The colored lines and symbols show the result of
    our simulations, quantitatively reproducing the observed
    correlation.  Colors and symbols are the same as in
    Figure~\ref{fig:amati}.  The Lorentz factors from the simulation
    are computed at the photosphere, since after the photons are
    released there is no internal pressure left to produce further
    acceleration of the ejecta.  The inset shows the Amati correlation
    highlighting in red the bursts for which the Lorentz factor could
    be measured.  Finally, smaller gray symbols show the Lorentz
    factor derived from observations under the assumption of a uniform
    external medium.}
\label{fig:gamma}}
\end{figure}

Figure~\ref{fig:gamma} shows a comparison between our results and the
observed correlation between the isotropic energy and the Lorentz
factor of the outflow (Liang et al. 2010; Ghirlanda et al. 2012a). The
observational correlation has significant scatter, possibly due to the
difficulty of precisely measuring the Lorentz factor of the
outflow. As an example of such difficulty we show the measurements
under the assumption of a wind environment with dark empty symbols and
the measurements under the assumption of uniform environment with a
solid gray symbol (from Ghirlanda et al. 2012). The two results are
indeed quite different. The Lorentz factor of the synthetic curves is
computed as the average of the Lorentz factor of the material at the
photospheric radius weighted with the local emissivity. Beyond the
photosphere the flow is not accelerated since the radiation decouples
and the outflow coasts until the external shock radius is reached,
when the observational Lorentz factor is measured.  Our simulations
reproduce the correlation for a wind environment both in slope and
normalization, but they have a smaller scatter. Simulated jets have,
however, lower Lorentz factor than the measured values for a uniform
environment. Our results extend to lower Lorentz factors with respect
to the measurements because it has not been possible to measure the
Lorentz factor of weak GRBs, so far (see the inset in
Figure~\ref{fig:gamma}). The agreement of the simulations with the
observations is entirely due to the jet-star interaction, since the
simulations adopt a constant $\Gamma_\infty=400$ input except for one
case, for which $\Gamma_\infty=100$ (the blue line and symbols). After
the interaction with the stellar material has taken place, the Lorentz
factor of the outflow at the photosphere is strongly dependent on the
polar angle, in such a way that the observed correlation is
reproduced.

\begin{figure}
\includegraphics[width=\columnwidth]{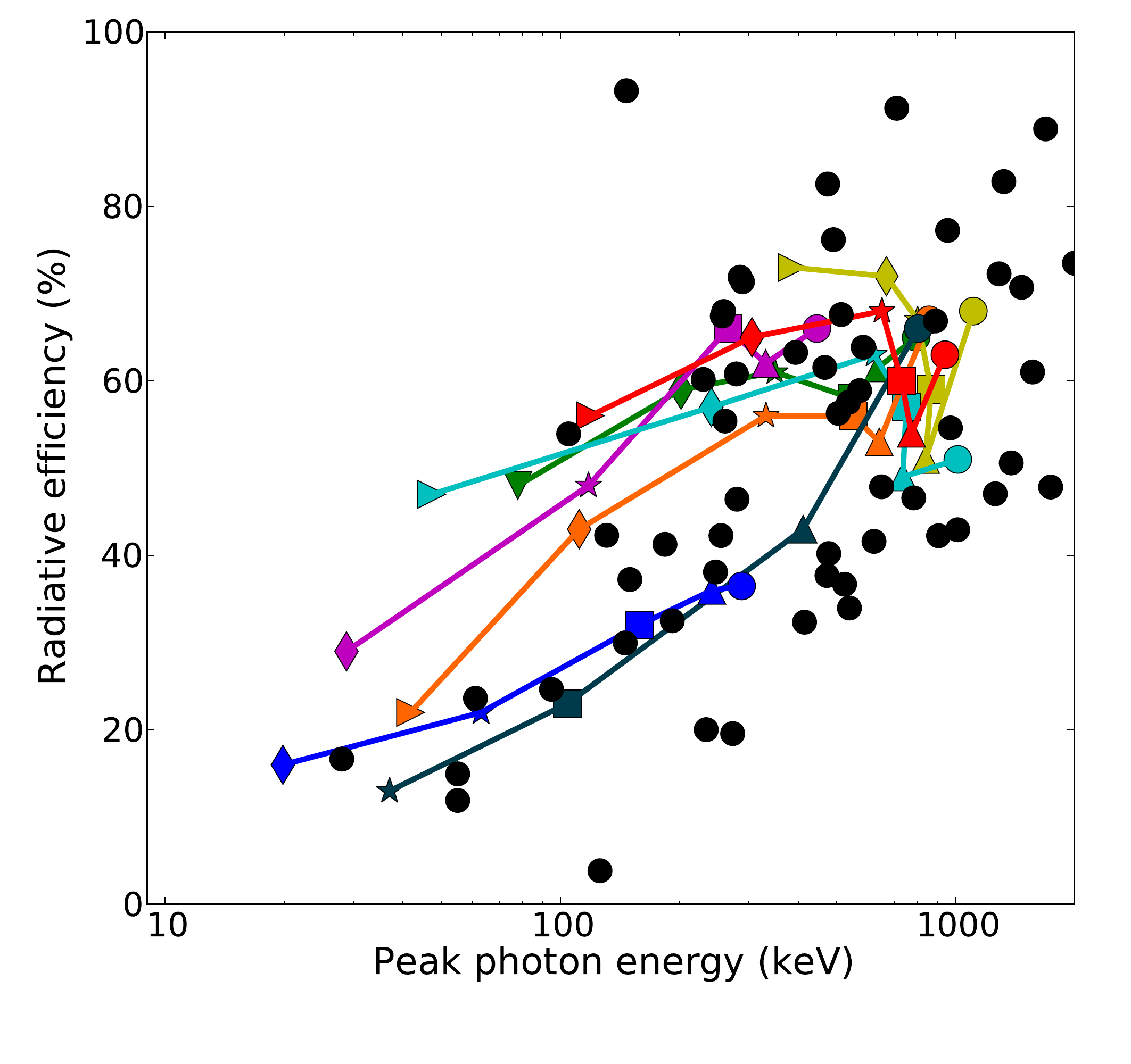}
\caption{{Local radiative efficiency of the photospheric emission,
    defined as the ratio of the energy released in photons to the
    total energy of the outflow.  A tendency of bursts with lower peak
    frequency (and therefore with lower isotropic-equivalent energy)
    to have lower efficiency is observed in the simulations. Solid
    black symbols show local radiative efficiencies
    estimated by comparing the prompt and early afterglow emission of
    observed long-duration GRBs.}
\label{fig:effi}}
\end{figure}

Finally, we can use our simulated light curves to compute the
radiative efficiency of the synthetic GRB prompt emission. We define
the radiation efficiency as the ratio of the energy released as
radiation at the photosphere over the total kinetic energy of the
flow. We find (see Figure~\ref{fig:effi}) that the simulations predict
a broad correlation between peak frequency and radiative efficiency
and, transitively, between the total energy radiated and the
efficiency.

Measuring the local radiative efficiency from observations is not an
easy task (see, e.g., Zhang et al. 2007), since the measurement must
be performed while the outflow is still relativistic in order to avoid
contributions from the ejecta moving at large angles with respect to
the line of sight. For this reason, radio calorimetry cannot be used.
The best proxy to the local radiative efficiency can be obtained,
under some assumptions, by comparing the brightness of the prompt
emission with that of the early X-ray afterglow. Within the framework
of an external shock dominated afterglow, the X-ray emission at early
times is produced by electrons in the cooling regime and is fairly
insensitive to the density of the ambient medium. Adopting the
analytic approximations of Panaitescu \& Kumar (2000) for a wind
environment with electrons in slow cooling, we can derive the kinetic
energy left in the fireball after the prompt emission as:
\begin{eqnarray}
E_k&=&10^{53}\left(\frac{17}{72}\right)^\frac{p}{p+2}10^{\frac{4}{2+p}(0.8p-59.63)}
\,L_\nu^{\frac{p}{2+p}}\,\left(\frac{\epsilon_e}{0.1}\right)^{4\frac{1-p}{2+p}}
\times \nonumber \\
&\times& \left(\frac{\epsilon_B}{0.01}\right)^\frac{2-p}{2+p}
\left(\frac{\nu}{10^{14.6}}\right)^\frac{2p}{2+p}
(1+z)^{-\frac{2p}{2+p}}
T_d^\frac{3p-2}{2+p}
\label{eq:ekin}
\end{eqnarray}
where $p$ is the slope of the non-thermal distribution of electrons,
$L_\nu$ is the luminosity density at a frequency $\nu$ above the
cooling frequency, $\epsilon_e$ and $\epsilon_B$ are the shock
equipartition parameters (i.e., the fraction of the downstream
internal energy used to accelerate relativistic electrons and to
generate or amplify the magnetic field, respectively), and $T_d$ is
the time in days after the GRB trigger. Note that Eq.~\ref{eq:ekin}
depends very weakly on $\epsilon_B$ for a reasonable value of the
electron acceleration slope $p\approx2$ and only the electron
equipartition parameter plays a dominant role in the resulting
calculation of the kinetic energy (Kumar \& Piran 2000).

In Figure~\ref{fig:effi} we show the observational efficiencies from a
sample of GRBs with X-ray afterglows measurements with solid black
symbols. We compute the $E_{X,\rm{iso}}$ released in the time interval
$(200,250)$~s, rest-frame using the best-fitting light-curves profiles
obtained by Margutti et al. (2013) in the common rest-frame energy
band 0.3-30 keV. Note that this estimate does not include the
contribution from X-ray flares. Only GRBs with early afterglow
re-pointing have been included and no extrapolation of the observed
temporal behavior has been performed. The resulting efficiencies
depend on the assumed $\epsilon_e$ value and therefore the agreement
observed in Figure~\ref{fig:effi} between the data and observations
should be seen as a consistency check rather than a proof. We adopted
fairly standerd values in the calculations ($\epsilon_e=0.1$,
$\epsilon_B=0.01$, $p=2.3$), but it has to be acknowledged that the
quantitative agreement would be lost for a smaller or bigger value of
$\epsilon_e$ (a qualitative good agreement is obtained for
$0.05\le\epsilon_e\le0.2$). It is worth mentioning, though, that the
general trend of higher efficiency for bursts with larger peak
frequency is robust and does not depend on the choice of
$\epsilon_e$. Finally it should be stressed that at the considered
time $(200-250)$~s there might still be some contribution to the X-ray
light curve form the prompt emission. We have also tried to use data
from an earlier ($100<t<150$~s) and a later ($500<t<750$~s) interval,
obtaining analogous results.

\section{Discussion and conclusions}

We presented the results of a set of hydrodynamic numerical
simulations of the evolution of long-duration GRB jets and their
progenitor stars. Thanks to the extended domain of the simulations we
were able to derive the location at which the spectral peak of the
advected radiation is set and use that information to compute light
curves from different progenitors and as seen by observers at
different off-axis angles with respect to the stars' rotation axis. We
find that within the assumptions made in this study, GRBs produced by
photospheric emission reproduce some observational correlations such
as the Amati correlation and the energy-Lorentz factor
correlation. The ability to account for such correlations, where other
radiation mechanisms have failed, suggests that the bulk of the GRB
prompt emission is released at the photosphere. Our results also
suggest that the primary driver of the difference between very
energetic events and weaker ones is the viewing angle, while the
scatter in the Amati correlation is due to different jet/progenitor
configurations.

Our simulation and the radiation code applied in post-processing do
not follow the full radiation transfer through the outflow and
therefore the radiation properties were computed assuming that the
spectrum of the entrained radiation is well described at all times by
a Planck function in thermal equilibrium with the leptons. In a
non-magnetized relativistic jet photon production mechanisms are
severely suppressed and the outflow becomes scattering-dominated well
before the photospheric radius (Beloborodov 2013; Vurm et
al. 2013). Under such conditions, especially in the presence of
dissipation, the radiation spectrum would be better approximated by a
Wien function with a harder $F(\nu)\propto\nu^3$ low-energy tail. We
note that, as long as the radiation and leptons are in thermal
equilibrium, the peak frequencies of the photons in a Planck and Wien
spectrum are equal to within a small factor $\nu_{\rm{peak,
    Planck}}/\nu_{\rm{peak, Wien}}=1.07$. Our results are therefore
only marginally affected by the choice of a Planck spectrum versus a
Wien one. We also note that a detailed discussion of the spectral
shape of the photospheric emission is beyond the scope of this paper
and we refer the reader to the extensive literature on the formation
of non-thermal spectra when sub-photospheric dissipation and/or
magnetic fields are present (Giannios 2006; Pe'er et al. 2006;
Giannios \& Spruit 2007; Beloborodov 2010; Lazzati \& Begelman 2010;
Vurm et al. 2011).

Another result of our simulations is a strong correlation between the
average Lorentz factor of the ejecta that contribute to the light
curve and the total radiated energy in the direction to the
observer. Such a correlation has been observed (Liang et al. 2010;
Ghirlanda et al. 2012) and the results of our simulations are in good
agreement with observations as long as a wind environment is assumed
in the calculation of the observational Lorentz factor. This is
another important piece of the puzzle, since the ``standard''
synchrotron internal shock model predicts the opposite correlation
(Ghisellini et al. 2010). Alternatively, the correlation might be
related to the entrainment of baryons from a neutrino-driven wind (Lei
et al. 2012). Finally, we found in our simulations that the radiative
efficiency of brighter bursts is higher than that of weaker bursts. By
comparing the early X-ray afterglow flux to the prompt emission energy
of a sample of well-observed bursts, we were able to show that the
correlation exists in the observations (see also Bernardini et
al. 2012). However, a detailed comparison requires the assumption of a
value for the electron equipartition parameter $\epsilon_e$, making
the quantitative comparison with the simulations
parameter-dependent. We show that adopting the fiducial value
$\epsilon_e=0.1$, we can obtain good agreement between the simulation
results and the observed values.

As a final note it should be remembered that the correlations that we
discuss here are not exhaustive of all correlations discovered in GRB
catalogs. In particular, there is speculation that the more physical
correlation is between the peak frequency and the luminosity, rather
than the energy release integrated over time (Yonetoku et
al. 2004). In addition, an analogous correlation with luminosity
exists within individual bursts, where the instantaneous peak
frequency correlates with the instantaneous luminosity (Lu et
al. 2012). Since our simulations assume, for lack of a better model,
an engine with constant luminosity, we cannot explore any correlation
that depends on the variability injected by the central engine.

\acknowledgements We thank S.E. Woosley and A. Heger for making their
pre-SN models available, Lorenzo Amati for sharing his up-to-date data
on the Amati correlation before publication, and Gabriele Ghisellini
and Dimitrios Giannios for valuable discussions. The software used in
this work was in part developed by the DOE-supported ASC/Alliance
Center for Astrophysical Thermonuclear Flashes at the University of
Chicago. This work was supported in part by the Fermi GI program
grants NNX10AP55G and NNX12AO74G (DL). BJM is supported by an NSF
Astronomy and Astrophysics Postdoctoral Fellowship under award
AST-1102796.

\end{document}